\documentclass[%
 reprint,
 superscriptaddress,
 amsmath,amssymb,
 aps,
 prb,
 longbibliography,
 floatfix
]{revtex4-2}

\usepackage{graphicx}

\usepackage[%
  colorlinks=true,
  urlcolor=blue,
  linkcolor=blue,
  citecolor=blue
]{hyperref}

\usepackage[bb=dsserif]{mathalpha}

\usepackage[normalem]{ulem}
\usepackage{braket}

\usepackage{pifont}

\let\Re\relax
\let\Im\relax

\DeclareMathOperator{\tr}{tr}
\DeclareMathOperator{\trn}{\bar{tr}}

\DeclareMathOperator{\Re}{Re}
\DeclareMathOperator{\Im}{Im}
\DeclareMathOperator{\even}{W_\mathrm{even}}
\DeclareMathOperator{\odd}{W_\mathrm{odd}}

\DeclareMathOperator{\atan}{atan2}

\newcommand{\logtau}{\overline{\mathsf{O}}^\mu}
\newcommand{\logOne}{\overline{\mathbb{1}}}
\newcommand{\logX}{\overline{X}}
\newcommand{\logY}{\overline{Y}}
\newcommand{\logZ}{\overline{Z}}

\newcommand{\thr}{\mathrm{th}}
\newcommand{\calM}{\hat{\mathcal{M}}_{\mu,s}}
\newcommand{\calH}{\hat{\mathcal{H}}_{\mu,s}}
\newcommand{\Hgate}{\hat{H}_{\mu,s}^{(l,m)}}
\newcommand{\Vgate}{\hat{V}_{\mu,s}^{(l,m)}}
\newcommand{\Tgate}{\hat{T}_j^{(\eta_j)}}
\newcommand{\calZ}{\mathcal{Z}_{\mu,s}}
\newcommand{\MWPM}{\text{(MWPM)}}
\newcommand{\inc}{\text{(inc)}}
\newcommand{\coh}{\text{(coh)}}

\def\tcm{T.C.M. Group, Cavendish Laboratory, University of Cambridge, J.J. Thomson Avenue, Cambridge, CB3 0HE, UK}
\def\DAMTP{DAMTP, University of Cambridge, Wilberforce Road, Cambridge, CB3 0WA, UK}

\begin{document}

\title{Statistical mechanical mapping and maximum-likelihood thresholds for the surface code under generic single-qubit coherent errors}

\author{Jan Behrends}
\affiliation{\tcm}

\author{Benjamin B\'eri}
\affiliation{\tcm}
\affiliation{\DAMTP}

\begin{abstract}
The surface code, one of the leading candidates for quantum error correction, is known to protect encoded quantum information against stochastic, i.e., incoherent errors. 
The protection against coherent errors, such as from unwanted gate rotations, is however understood only for special cases, such as rotations about the $X$ or $Z$ axes.
Here we consider generic single-qubit coherent errors in the surface code, i.e.,  rotations by angle $\alpha$ about an axis that can be chosen arbitrarily. 
We develop a statistical mechanical mapping for such errors and perform  entanglement analysis in transfer matrix space to numerically establish the existence of an error-correcting phase, which we chart in a subspace of rotation axes to estimate the corresponding maximum-likelihood thresholds $\alpha_\thr$. 
The classical statistical mechanics model we derive is a random-bond Ising model with complex couplings and four-spin interactions (i.e., a complex-coupled Ashkin-Teller model).
The error correcting phase, $\alpha<\alpha_\thr$, where the logical error rate decreases exponentially with code distance, is shown to correspond in transfer matrix space to a gapped one-dimensional quantum Hamiltonian exhibiting spontaneous breaking of a $\mathbb{Z}_2$ symmetry.  
Our numerical results rest on two key ingredients: (i) we show that the state evolution under the transfer matrix---a non-unitary (1+1)-dimensional quantum circuit---can be efficiently numerically simulated using matrix product states.
Based on this approach, (ii) we also develop an algorithm to (approximately) sample syndromes based on their Born probability. 
The $\alpha_\thr$ values we find show that the maximum likelihood thresholds for coherent errors are larger than those for the corresponding incoherent errors (from the Pauli twirl), and significantly exceed the values found using minimum weight perfect matching.
\end{abstract}

\maketitle

\section{Introduction}

Quantum error correction (QEC) is a crucial step for quantum computers to achieve quantum advantage~\cite{Shor:1995fj,Calderbank:1996ja,Steane:1996kg}.
While random circuit sampling~\cite{Arute:2019fg,Wu:2021em,Madsen:2022jm} and Hamiltonian dynamics~\cite{Kim:2023bm} have been demonstrated on quantum computers, significantly reducing noise is believed to be necessary for these systems to not be classically efficiently simulable~\cite{Dalzell2021,StilckFranca:2021cb,Hangleiter:2023cq}.
QEC can achieve such a significant suppression of noise~\cite{Preskill:2018gt}.
As a first step towards this goal, recent experiments demonstrated the feasibility of QEC in superconducting circuits~\cite{Krinner:2022dk,Acharya:2023fl,Acharya:2025cn} and cold atom arrays~\cite{Bluvstein:2024ht} by implementing the surface code~\cite{Bravyi1998quantum,Freedman1998projective} and the closely related color code~\cite{Bombin:2006hw}.

QEC is usually viewed as combating decoherence.
Decoherence~\cite{Zurek:2003fm} can be effectively described by purely stochastic noise that acts with a certain probability, i.e., incoherently, on a qubit~\cite{Dennis:2002ds,Bravyi:2018ea}.
However, quantum errors may also arise because of other processes:
Since time evolution of qubits is unitary, any unwanted time evolution leads to unwanted rotations of qubits.
These so-called coherent errors may thus arise, e.g., due to erroneous quantum gates or their imperfect implementation.

Coherent errors may pose a challenge for QEC because they can add up constructively by quantum inference~\cite{Greenbaum:2018ce,Bravyi:2018ea,Gottesman2019,Iverson:2020fe} and thus accumulate faster than incoherent, i.e., probabilistic, errors~\cite{Wallman:2014hs,Wallman:2016be,Gottesman2019}.
Interference effects also make it challenging to address the impact of coherent errors on QEC codes~\cite{Iverson:2020fe}.
Error mitigation~\cite{Chamberland:2017dm,Cai:2020bz}, e.g., via randomized compiling~\cite{Kern:2005gq,Wallman:2016be,Hashim:2021gr,Winick2022}, can facilitate QEC by converting coherent into well-understood incoherent errors~\cite{Wallman:2016be}.
However, these mitigation techniques require additional resources, e.g., randomization over several cycles~\cite{Wallman:2016be}, which makes it desirable to avoid error mitigation techniques if possible.
This raises the more fundamental question about coherent errors in QEC:
How does their impact scale with code distance?
Is there a fundamental error threshold below which coherent errors can be corrected in sufficiently large systems?

A physically illuminating approach to studying these questions is though statistical mechanics mappings~\cite{Dennis:2002ds,Katzgraber:2009iz,Venn:2023fp,Wille:2024ks}.
Such mappings have established thresholds for incoherent errors~\cite{Dennis:2002ds,Katzgraber:2009iz}, including depolarizing noise and generic incoherent uncorrelated single-qubit errors~\cite{Bombin:2012km,Wootton:2012cb,Chubb:2021cn}.
For coherent $X$ or $Z$ rotations, a similar statistical mechanics mapping demonstrates a surprisingly large error threshold~\cite{Venn:2023fp}, explaining why standard decoders can achieve high thresholds without error mitigation~\cite{Bravyi:2018ea,Venn:2020ge,Marton:2023da}.
Generic single-qubit errors are however largely unexplored: It is unknown whether results for coherent $X$ (or $Z$) rotations generalize for generic rotation axes or whether even an error threshold exists.
Tensor-network simulations~\cite{Darmawan:2017fx} can be used to study those generic coherent errors, including their impact on teleportation protocols~\cite{Eckstein:2024ev}, but so far results exist only for moderate system sizes~\cite{Darmawan2024}.

\begin{figure*}
\includegraphics[scale=1]{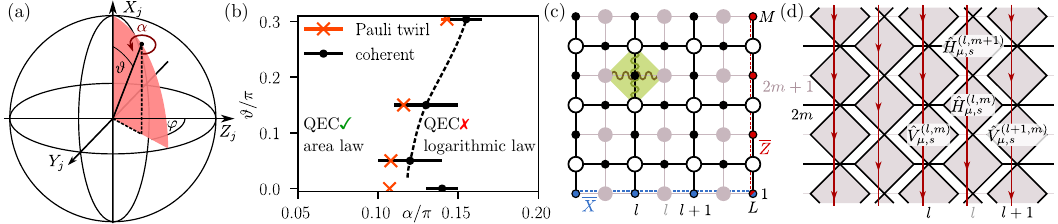}
\caption{Summary of results.
(a) Bloch sphere visualizing the coherent error from a unitary rotation of single qubits [Eq.~\eqref{eq:rotation}].
We numerically simulate rotations by the angle $\alpha$ about three different axes in the red shaded plane with $\varphi=\pi/4$ and $\vartheta = 0.05\pi,0.15\pi,0.304\pi$.
(b) Approximate phase diagram for coherent errors and $\varphi=\pi/4$. In the phase with $\alpha$ below threshold, the logical error rate decays exponentially with distance (QEC\ding{51}); the dual quantum circuit yields a 1D area law for entanglement.
Above threshold, the logical error rate decays slowly (QEC\ding{55}) and the dual quantum circuit yields a logarithmic entanglement phase. Black dots denote conservative threshold estimates based on maximal entanglement fluctuations~\cite{Kjall:2014bd}, and the error bars arise from the uncertainty of this maximum. The circuit is Gaussian at $\vartheta=0$, which enables the simulation of larger system sizes, yielding a more accurate threshold (value at $\vartheta=0$ from Ref.~\onlinecite{Venn:2023fp}).
The orange crosses mark the corresponding incoherent Pauli twirl~\cite{Emerson:2007ej,Silva:2008ij} threshold~\cite{Dennis:2002ds,Bombin:2012km}.
(c) The surface code geometry we study, illustrated for $L=M=4$.
The direct lattice is shown in black, with physical qubits on its links (black dots), stabilizers $S_v^X$ (white disks) on its vertices, and $S_p^Z$ (gray disks) on its faces; the dual lattice is shown in gray.
The blue dashed line shows a representative for the logical $\logX$ and the red dashed line one for the logical $\logZ$.
We map the surface code to a RBIM (Sec.~\ref{sec:stat_mech}) with complex vertex-vertex couplings $J^x$ (wiggly light green line), plaquette-plaquette couplings $J^z$ (wiggly dark green line), and four-spin interactions $J^y$ (light green square) that couple direct and dual lattice.
(d) Transfer matrix representation (Sec.~\ref{sec:quantum_circuit}): The three coupling terms act as three-qubit gates $\Hgate$ and $\Vgate$.
To sample error strings (Sec.~\ref{sec:algorithm}), for each site we successively (visualized by the red line) apply one of four possible three-qubit gates, sampled based on its probability.
}
\label{fig:layout}
\end{figure*}

In this work, we develop a statistical mechanics mapping~\cite{Dennis:2002ds,Katzgraber:2009iz,Venn:2023fp,Wille:2024ks,Bombin:2012km,Wootton:2012cb,Chubb:2021cn} to obtain maximum likelihood thresholds for the surface code under generic single-qubit coherent errors, i.e., unitary rotations about any axis, assuming perfect syndrome measurements.
We focus on the surface code since it is one of the most promising experimentally relevant candidates~\cite{Krinner:2022dk,Acharya:2023fl,Bluvstein:2024ht} for QEC thanks to its local connectivity and large incoherent error thresholds~\cite{Dennis:2002ds,Venn:2023fp}.
We describe an algorithm based on our statistical mechanics mapping that efficiently samples error strings. For coherent errors, this was previously possible only for $X$ (or $Z$) rotations~\cite{Bravyi:2018ea,Venn:2023fp} or non-Pauli errors close to the incoherent limit~\cite{Ma:2023ia}.
Our results facilitate computing both maximum likelihood and decoder-dependent coherent-error thresholds~\cite{Bravyi:2018ea,Venn:2020ge,Marton:2023da}.

We show that the maximum likelihood threshold is larger than the one obtained for incoherent errors of the same magnitude (i.e., under the so-called Pauli twirl approximation~\cite{Emerson:2007ej,Silva:2008ij}), despite the potential constructive inference of coherent errors~\cite{Greenbaum:2018ce,Bravyi:2018ea,Gottesman2019,Iverson:2020fe}.
The above-threshold phase that we find is markedly distinct from that for incoherent errors~\cite{Dennis:2002ds,Merz:2002gj,Bombin:2012km}.
It can be interpreted as a critical phase characterized by a slow decay of the logical error with system size.

\subsection{Overview of the Results}
\label{sec:results}

Before describing the technical details, we give an overview of our approach and results.
A pictorial summary is shown in Fig.~\ref{fig:layout}.

For local unitary rotations about any axis, we develop a statistical mechanics mapping~\cite{Dennis:2002ds,Katzgraber:2009iz,Venn:2023fp,Wille:2024ks,Bombin:2012km,Wootton:2012cb,Chubb:2021cn} that expresses error amplitudes as complex-coupling (``complex'' henceforth) partition functions.
Components of the rotation about the $X$ and $Z$ axes map to complex two-dimensional (2D) random bond Ising models (RBIM) on the direct and dual lattice, respectively.
Similarly to incoherent errors~\cite{Bombin:2012km,Wootton:2012cb,Chubb:2021cn}, rotation components about the $Y$ axis give four-spin interactions that couple the direct and dual lattice.
The resulting complex two-layer RBIM with inter-layer interactions is variant of the Ashkin-Teller model~\cite{Ashkin:1943jd} with disordered complex couplings, i.e., a complex variant of the eight-vertex model~\cite{Baxter:1971fk}.

Due to its complex couplings, this RBIM suffers from a sign problem and thus cannot be solved by Monte Carlo simulations.
Instead, we simulate its transfer matrix using matrix product states (MPSs)~\cite{Hauschild:2018bp,Cirac:2021gx}.
MPS simulations are feasible since, as we demonstrate, states in the transfer matrix space exhibit an area law below threshold.
This substantially expands the scope of this feature, previously observed only for $X$ (or $Z$) rotations~\cite{Behrends:2024bs}.
(Similar MPS simulations have been used to address depolarizing noise~\cite{Bravyi:2014ja}.)
Above threshold, the entanglement entropy grows with system size, again demonstrating the general scope of this feature beyond the special case of $X$ (or $Z$) rotations~\cite{Behrends:2024bs}.
This is distinct from the above-threshold phase of incoherent errors, where the entanglement entropy follows an area law~\cite{Behrends:2024bs}.

To use our statistical mechanics approach for estimating code performance, we must sample syndromes. For coherent errors, this cannot be done straightforwardly since different error strings corresponding to the same syndrome generally interfere~\cite{Iverson:2020fe}.
Based on the transfer matrix representation, we develop an algorithm that samples syndromes according to their probability, in the setting relevant to computing error rates captured by average infidelities.
We achieve this by sampling bonds of the statistical mechanics model, which is equivalent to sampling equivalence classes of error strings yielding the same syndrome. 
The scope of our approach is substantially beyond that of algorithms tailor-made for the special case of $X$ (or $Z$) rotations~\cite{Bravyi:2018ea}, or schemes for non-Pauli error channels are close to the incoherent limit~\cite{Ma:2023ia}.

We numerically investigate three different rotation axes as shown in Fig.~\ref{fig:layout}(a).
We find two distinct phases of QEC separated by a threshold, which we show in the phase diagram in Fig.~\ref{fig:layout}(b):
Below threshold, the minimum infidelity between the encoded and post-QEC state (with the minimum with respect to the possible Pauli corrections) decreases exponentially with code distance.
Above threshold, the minimum infidelity continues to decrease with code distance, but much more slowly than exponential.
This above-threshold phase is markedly different from the corresponding incoherent-error phase, where the logical error rate increases with system size~\cite{Dennis:2002ds,Bombin:2012km,Wootton:2012cb,Chubb:2021cn,Bravyi:2014ja}.
The threshold angles for maximum likelihood decoding coherent errors are larger than that for their Pauli twirl approximation [Fig.~\ref{fig:layout}(b)], which replaces the coherent error channel by a corresponding incoherent channel via averaging over the Pauli group~\cite{Emerson:2007ej,Silva:2008ij}.

We also compare the thresholds from our (MPS approximation of the) maximum-likelihood decoder with those for the much faster minimum weight perfect matching (MWPM) decoder~\cite{Kolmogorov:2009dk,Fowler:2012ji,Fowler:2015hb,pymatchingv1}. For incoherent errors, the MWPM threshold is known to be close to the maximum-likelihood bound~\cite{Dennis:2002ds,Bombin:2012km,Wootton:2012cb,Chubb:2021cn,Bravyi:2014ja}.
We show that for coherent errors the MWPM thresholds are smaller than both Pauli twirl and maximum likelihood thresholds.

The remainder of this work is organized as follows:
After introducing coherent errors in Sec.~\ref{sec:coherent}, we map these errors to a complex four-spin RBIM in Sec.~\ref{sec:stat_mech} and subsequently to a quantum circuit in Sec.~\ref{sec:quantum_circuit}.
We then describe in Sec.~\ref{sec:algorithm} how the circuit can be utilized for sampling error strings, and present numerical results including error thresholds in Sec.~\ref{sec:numerics} before concluding in Sec.~\ref{sec:conclusion}.

\section{Surface Code and Error Models}
\label{sec:coherent}

Surface codes are topological stabilizer codes~\cite{Gottesman97,Bravyi1998quantum,Kitaev:2003jw,Fowler:2012ji,Terhal:2015ks} whose logical subspace is the $+1$ eigenspace of mutually commuting vertex $S_v^X = \prod_{j\in v} X_j$ and plaquette stabilizers $S_p^Z  =\prod_{j \in p} Z_j$, where $X_j$ and $Z_j$ are physical qubit operators on links of a square lattice; cf.\ Fig.~\ref{fig:layout}(c).
We consider a planar square lattice with smooth edges at left and right sides, and rough edges at top and bottom~\cite{Bravyi1998quantum}.
Apart from the boundaries, vertex and plaquette operators act on four neighboring qubits each.
In this geometry, the stabilizer $+1$ eigenspace encodes one logical qubit: The logical $\bar{X} = \prod_{j\in \gamma} X_j$ connecting left and right boundaries, and logical $\bar{Z} = \prod_{j\in \gamma'} Z_j$ connecting top and bottom boundaries commute with all stabilizers (i.e., act within the logical subspace), but mutually anticommute since they overlap on an odd number of sites; we choose the logical operators as shown in Fig.~\ref{fig:layout}(c) and denote $\bar{X}$'s path length by $L$ and $\bar{Z}$'s path length by $M$.
The total number of qubits $N=L M +(L-1)(M-1)$.

Starting from an initial logical state $\ket{\psi}$, we consider coherent errors that change $\ket{\psi}\to U \ket{\psi}$, where the unitary operator
\begin{align}
 U=\prod_j U_j , & & U_j = \exp \left(i \alpha_j \mathbf{n}_j \cdot \boldsymbol{\mathsf{O}}_j \right)
 \label{eq:rotation}
\end{align}
is the product of local unitary rotations by an angle $\alpha_j$ about the axis $\mathbf{n}_j = (\cos \vartheta_j,\sin \vartheta_j \sin \varphi_j, \sin \vartheta_j \cos \varphi_j)$; here $\boldsymbol{\mathsf{O}}_j = (X_j,Y_j,Z_j)$ is the vector of physical qubit operators at site $j$.
Expressed as an error channel, coherent errors change an initial state $\rho = \ket{\psi}\bra{\psi}$ as
\begin{equation}
 \mathcal{E} [\rho] = U \rho U^\dagger .
 \label{eq:error_channel}
\end{equation}

By contrast, incoherent errors act probabilistically.
The corresponding local error channel~\cite{Dennis:2002ds,Bombin:2012km,Chubb:2021cn}
\begin{equation}
 \mathcal{E}_j^\inc [\rho] = (1-p_j) \rho + \sum_{k=1}^3 p_{j,k} \mathsf{O}_j^k \rho \mathsf{O}_j^k
 \label{eq:incoherent_channel}
\end{equation}
with $p_j = \sum_{k=1}^3 p_{j,k}$ changes a state $\ket{\psi} \to \mathsf{O}_j^k \ket{\psi}$ with probability $p_{j,k}$, or leaves it invariant with probability $1-p_j$.
The full error channel $\mathcal{E}^\inc = \bigotimes \mathcal{E}_j^\inc$ thus changes an initial pure state into a probabilistic mixture of states, in contrast to purely coherent errors $\mathcal{E} [\rho]$ that transform pure states into pure states via a unitary transformation.
We will contrast some of our findings for coherent errors with the incoherent Pauli twirl approximation~\cite{Emerson:2007ej,Silva:2008ij}, which replaces each local unitary rotation $U_j$ by the incoherent channel
\begin{equation}
 U_j \rho U_j^\dagger \to \mathcal{E}_j^\inc [\rho]
 \label{eq:pauli_twirl}
\end{equation}
with error probabilities $p_{j,k} = n_{j,k}^2 \sin^2 \alpha_j$ ($k=1,2,3$), where $n_{j,k}$ are the components of $\mathbf{n}_j$ in Eq.~\eqref{eq:rotation}.
In this approximation, all contributions in Eq.~\eqref{eq:error_channel} that are of the form $\mathsf{O}_j^\mu \rho \mathsf{O}_j^\nu$ with $\mu\neq \nu$ are neglected.

\subsection{QEC for the Surface Code}
\label{sec:qec_within_surface_code}

QEC consists of two steps: After first measuring all stabilizers given the post-error state $U\ket{\psi}$, one applies a Pauli correction operation that returns the state to the logical subspace.
Throughout this work, we assume that all measurements and corrections are perfect.
Measuring all stabilizers projects the post-error state onto a measurement outcome characterized by a syndrome $s$ that is defined by the positions of all $S_p^Z=-1$ and $S_v^X=-1$.
The post-measurement state can be brought back to the logical subspace by applying a Pauli correction $C_s$ consisting of $X_j$ and $Z_j$ operators that connect pairs of $S_p^Z=-1$ and $S_v^X=-1$, respectively.
The string $C_s$ is not unique:
First, a syndrome $s$ can also be corrected by $C_{s}'$ that equals $C_s$ up to closed loops of Pauli $X_j$-strings and Pauli $Z_j$-strings because, as closed loops are products of stabilizer operators, both strings correspond to the same syndrome.
Second, one can correct $s$ by $\logtau C_s$, the original Pauli string times a logical operator $\logtau \in \{\logOne,\logX,\logY,\logZ\}$ for $\mu \in \{0,1,2,3\}$.
While the state $C_s'\ket{\psi}=C_s\ket{\psi}$ is invariant upon adding closed loops (since $\ket{\psi}$ is a $+1$ eigenstate of all stabilizers), changing $C_s \to \logtau C_s$ is equivalent to acting with a logical operator on $\ket{\psi}$.
Thus, after syndrome measurement, one needs to select one of the four $\logtau C_s$ that yields a state closest to the initial state, which in practice is done by a decoder.
Here, we aim for the decoder-independent theoretical optimum, i.e., maximum likelihood decoding by one of the four $\logtau C_s$.

\subsection{Logical Error Rate}
\label{sec:logical_error_rate}

The probability of observing a syndrome $s$ upon syndrome measurement on the state $\mathcal{E}[\rho]$ in general depends on $\rho$.
Hence it should be viewed as a conditional probability $P(s|\rho)$.
This $\rho$ dependence arises also in our setup [Fig.~\ref{fig:layout}(b)].
($P(s|\rho)$ would be $\rho$-independent only if all stabilizers had even weight~\cite{Bravyi:2018ea}---i.e., act on an even number of qubits---which is not true in our setup due to the boundaries.)
To mitigate the dependence on the initial state, we average all quantities of interest over the Bloch sphere, assuming a uniform distribution of $\rho$, i.e., we take the average $\langle \dots \rangle_\Omega = \int_\Omega d\rho [\dots] P(\rho|s)$, where we use $P(\rho|s)$ because we average over $\rho$ for a given syndrome $s$~\cite{Venn:2020ge}.

The Bloch-sphere averaged infidelity~\cite{Fuchs:1999ce,Wallman:2015il,Sanders:2015iw} captures the decoder-independent feasibility of QEC.
To construct this QEC measure, consider the fidelity
\begin{equation}
  F(\rho, \rho_{\mu,s} ) = \frac{\braket{\psi|\rho_{\mu,s} |\psi}}{\tr[\rho_{\mu,s}]} = \frac{|\braket{\psi| \logtau D_s | \psi}|^2}{P(s|\rho)} 
  \label{eq:fidelity}
\end{equation}
between an initial logical state $\rho=\ket{\psi}\bra{\psi}$ and $\rho_{\mu,s} = \logtau D_s \rho D_s^\dagger \logtau$ with
\begin{equation}
 D_s = \Pi_0 C_s U \Pi_0
 \label{eq:Ds_definition}
\end{equation}
where $\Pi_0$ projects onto the logical subspace.
The state $\rho_{\mu,s}$ is the post-error and post-measurement state after applying the correction $\logtau C_s$, which is not normalized, but satisfies
\begin{equation}
P(s|\rho) = \tr[\Pi_s U\rho U^\dagger] = \tr[C_s^\dagger \Pi_0 C_s U\rho U^\dagger] = \tr[\rho_{\mu,s}],
\end{equation}
where $\tr[\rho_{\mu,s}]$ is $\mu$-independent by $(\logtau)^2 = \mathbb{1}$.
As we now show, the numerator of Eq.~\eqref{eq:fidelity} is, up to a factor of $2$, the probability of the error $\logtau C_s$ or equivalent errors (i.e., equal to $\logtau C_s$ times a stabilizer product):
Using $\Pi_0 = (1/2) \sum_{\mu=0}^3 \logtau \rho \logtau$, we split the conditional probability
\begin{equation}
 P(s|\rho) = \tr[D_s^\dagger D_s \rho] = \frac{1}{2} \sum_{\mu=0}^3 | \braket{\psi|\logtau D_s|\psi}|^2
\end{equation}
into four parts that equal the numerator of $F(\rho,\rho_{\mu,s})$ for each of $\mu \in \{ 0,1,2,3 \}$.
The fidelity can thus be interpreted, up to a factor of $1/2$, as the probability of the error string $\logtau C_s$ (or equivalent strings), normalized by the corresponding syndrome probability $P(s|\rho)$.
Maximum likelihood decoding means choosing, given a syndrome $s$, the correction operation $\logtau C_s$ with the largest $F(\rho,\rho_{\mu,s})$.
Error correction is successful when the chosen $\logtau C_s$ returns the error-corrupted state to a state that approaches the initial state exponentially with code distance.

We now take the Bloch-sphere average over the infidelity (i.e., one minus the fidelity) $r_{\mu,s} = \langle 1 - F(\rho,\rho_{\mu,s} ) \rangle_\Omega$.
Minimized over all logical operators, the scaling of the minimum infidelity $\min_\mu r_{\mu,s}$ with code distance is a proxy for the feasibility of QEC:
When the minimum infidelity goes to zero, one error string $\logtau C_s$ brings the post-error state to a state close to the initial state, i.e., QEC is possible.
When the minimum approaches a nonzero constant (in the worst-case scenario $1/4$, corresponding all options having the same overlap squared with the initial state), QEC is not possible.

The diamond-norm distance $D_\diamondsuit$~\cite{Sanders:2015iw,Wallman:2014hs} is a worst-case measure for QEC as it is maximized with respect to all possible pure initial states.
Since $D_\diamondsuit$ scales maximally as $\sqrt{r_{\mu,s}}$~\cite{Wallman:2016be,Gottesman2019,Iverson:2020fe}, we can infer the worst-case scaling of the diamond-norm distance from the scaling of $r_{\mu,s}$ with code distance.
Throughout this work, we use as the logical error rate the minimum Bloch-sphere averaged infidelity averaged over syndromes
\begin{equation}
 P_L = \langle \min_\mu r_{\mu,s} \rangle_s ,
 \label{eq:logical_error}
\end{equation}
where $\langle \dots \rangle_s$ denotes the syndrome average with respect to $P(s) = \int_\Omega d\rho P(s|\rho) P(\rho)$ with uniform $P(\rho)=1/(4\pi)$.

Due to the projection onto the logical subspace in Eq.~\eqref{eq:Ds_definition}, $D_s$ must be of the form
\begin{equation}
 D_s = \mathcal{Z}_{0,s} \logOne +  \mathcal{Z}_{1,s} \logX + \mathcal{Z}_{2,s} \logY + \mathcal{Z}_{3,s} \logZ ;
 \label{eq:Ds_coefficient}
\end{equation}
no operations that leave the logical subspace are possible.
As we show next in Sec.~\ref{sec:stat_mech}, each coefficient $\mathcal{Z}_{\mu,s} = \tr[ D_s \logtau]$ can be expressed as the partition function of two interacting 2D complex RBIM.
The average infidelity equals
\begin{equation}
 r_{\mu,s} = \frac{2}{3} \frac{ \sum_{\nu \neq \mu} | \mathcal{Z}_{\nu,s}|^2}{P(s)},
 \label{eq:average_infidelity}
\end{equation}
where $P(s) = \sum_\nu |\mathcal{Z}_{\nu,s}|^2$ is the Bloch-sphere averaged syndrome probability, as we show in Appendix~\ref{sec:infidelity}.

\section{Statistical Physics Mapping}
\label{sec:stat_mech}

We now introduce the statistical mechanics mapping to express the coefficients $\mathcal{Z}_{\mu,s}$ as complex random bond Ising model partition functions.
The partition function arises from expressing the sum over closed loop configurations, which arise when enumerating errors that differ from each other by stabilizer products, as a sum over configurations of classical Ising spins.

\begin{table*}
\caption{\label{tab:phases}
Set of imaginary parts of the couplings:
The entries denote the set $\{ \Im J_j^0 , \Im J_j^x, \Im J_j^y, \Im J_j^z \}$. This choice not unique.}
\begin{ruledtabular}
\begin{tabular}{l|cccc}
  & $0 < \varphi_j<\pi/2$ & $\pi/2 < \varphi_j<\pi$ & $\pi < \varphi_j<3\pi/2$ & $3\pi/2 < \varphi_j< 2\pi$ \\
  \hline
  $0 < \vartheta_j <\pi/2$ & $\{ \pi/2,-\pi/4,0,-\pi/4 \}$ & $\{ -\pi/4,\pi/2,-\pi/4,0 \}$ & $\{ -\pi/2, \pi/4,0,\pi/4 \}$ & $\{ \pi/4, -\pi/2,\pi/4, 0 \}$ \\
  $\pi/2 < \vartheta_j <\pi$ & $\{ \pi/4,0,-\pi/4,0 \}$ & $\{ 0,\pi/4,0,-\pi/4 \}$ & $\{ -\pi/4, 0, \pi/4,0 \}$ & $\{ 0,-\pi/4,0,\pi/4 \}$
\end{tabular}
\end{ruledtabular}
\end{table*}

\begin{figure}[b]
\includegraphics[scale=1]{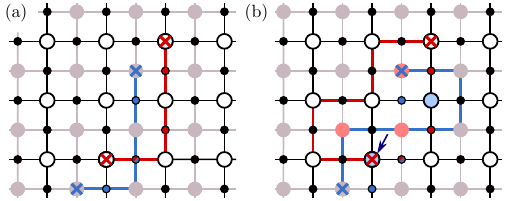}
\caption{(a) Example of a syndrome that is characterized by $S_v^X =-1$ and $S_p^Z=-1$, marked by red and blue crosses, respectively (the white and gray disks without crosses denote $S_v^X=+1$ and $S_p^Z=+1$ eigenvalues).
A reference string $\logtau C_s$ connects pairs of crosses of the same type: The blue and red lines denote Pauli strings of $X_j$ operators and $Z_j$ operators, respectively.
In the RBIM, reference strings translate to signs $\eta_{\mu,s}^{x,j}=-1$ for sites with $X_j$ (blue), and $\eta_{\mu,s}^{z,j}=-1$ for sites with $Z_j$ (red); cf.\ Eq.~\eqref{eq:reference_string}.
(b) Pauli string $P_\varsigma$ that equals $\logtau C_s$ up to closed loops.
In the RBIM, closed loops that include $S_v^X$ and $S_p^Z$ translate to the spins $\sigma_v=-1$ (blue vertices) and $\tilde{\sigma}_p=-1$ (red plaquettes), respectively.
In this example, the Pauli strings $\logtau C_s$ [panel (a)] and $P_\varsigma$ [panel (b)] anticommute:
An odd number of vertices [here: one vertex marked by a dark blue arrow in (b)] is simultaneously a $S_v^x=-1$ syndrome and part of a loop characterized by $\sigma_v=-1$.}
\label{fig:error_strings}
\end{figure}

We start by expanding each local unitary [Eq.~\eqref{eq:local_unitary}] as a sum of Pauli strings, using the parametrization
\begin{equation}
 U_j = \sum_{\substack{ \mathsf{x}_{j}=\pm1 \\ \mathsf{z}_j=\pm1}} e^{J_j^0 + \mathsf{x}_{j} J_j^x + \mathsf{x}_{j} \mathsf{z}_{j} J_j^y + \mathsf{z}_{j} J_j^z} X_j^{(1-\mathsf{x}_{j})/2}  Z_j^{(1-\mathsf{z}_{j})/2}
 \label{eq:local_unitary}
\end{equation}
with the complex couplings $J_j^\mu$ defined via
\begin{subequations}\begin{align}
e^{J_j^0 + J_j^x + J_j^y + J_j^z} &= \cos\alpha_j \\
e^{J_j^0 - J_j^x - J_j^y + J_j^z} &= i\sin\alpha_j \cos \vartheta_j \\
e^{J_j^0 - J_j^x + J_j^y - J_j^z} &= -\sin\alpha_j \sin \vartheta_j \sin \varphi_j \\
e^{J_j^0 + J_j^x - J_j^y - J_j^z} &= i\sin\alpha_j \sin \vartheta_j \cos \varphi_j .
\end{align}\label{eq:Jjmu_def}\end{subequations}
These relations are sufficient to define the couplings $J_j^\mu$ as they enter calculations only in exponentials.
For completeness, we give consistent choices for $\Im J_j^\mu$ (that are not unambiguously defined due to the periodicity of the complex exponential) in each octant of the Bloch sphere in Table~\ref{tab:phases}.
Using this parametrization, the rotation
\begin{equation}
 U= \prod_j U_j = \sum_{\varsigma}  c_{\varsigma} P_\varsigma
 \label{eq:U_pauli_strings}
\end{equation}
can be expanded as a sum of Pauli strings $P_\varsigma$ with coefficient $c_\varsigma$ labeled by configurations $\varsigma = \{\mathsf{x}_{j},\mathsf{z}_{j}\}$.
We express the coefficients of $D_s$ [Eq.~\eqref{eq:Ds_coefficient}] with Eq.~\eqref{eq:U_pauli_strings} as a sum of Pauli strings
\begin{align}
 \mathcal{Z}_{\mu,s} = \trn [\logtau D_s] = \sum_\varsigma c_\varsigma \trn[ \logtau C_s P_\varsigma \Pi_0 ],
 \label{eq:zpartition}
\end{align}
where $\trn$ denotes the trace normalized by the Hilbert space dimension, i.e., $\trn [\mathbf{1}_n] = 1$.
The projector onto the logical subspace
\begin{equation}
\Pi_0 = \sum_{\{ n_v, \tilde{n}_p \}} \prod_p (S_p^Z)^{\tilde{n}_p} \prod_v (S_v^X)^{n_v}
\end{equation}
equals the sum over all closed loops of Pauli strings, where we denote configurations of $n_v\in\{0,1\}$ and $\tilde{n}_p\in \{0,1\}$ by $\{n_v,\tilde{n}_p \}$.
Thus, the trace $\trn[ \logtau C_s P_\varsigma \Pi_0 ]\neq 0$ only when $P_\varsigma$ equals $\logtau C_s$ up to such a closed-loop configuration and a sign~\footnote{The strings can differ only by a sign (and not a general phase) since all operators are Pauli strings of only $X$ and $Z$ operators.}, i.e., when
\begin{equation}
P_\varsigma = \pm \logtau C_s \prod_{p,v} (S_p^Z)^{\tilde{n}_p} (S_v^X)^{n_v}
\end{equation}
for some set of $n_v$ and $\tilde{n}_p$.
Equating $X_j$ and $Z_j$ Pauli strings fixes $\mathsf{x}_j = \eta_{\mu,s}^{x,j} \sigma_{v_j} \sigma_{v_j'}$ and $\mathsf{z}_j = \eta_{\mu,s}^{z,j} \tilde{\sigma}_{p_j} \tilde{\sigma}_{p_j'}$, where $v_j^{(\prime)}$ and $p_j^{(\prime)}$ label the two vertices and plaquettes connected by the qubit $j$, the configurations of classical Ising spins $\sigma_v = (-1)^{n_v}$ and $\tilde{\sigma}_p  = (-1)^{\tilde{n}_p}$ label the closed-loop configurations, and the signs $\eta_{\mu,s}^{x,j} = \pm 1$, $\eta_{\mu,s}^{z,j} = \pm1$ define the reference string
\begin{equation}
 \logtau C_s =\left( \prod_j Z_j^{(1-\eta_{\mu,s}^{z,j})/2} \right) \left( \prod_j X_j^{(1-\eta_{\mu,s}^{x,j})/2} \right)
 \label{eq:reference_string}
\end{equation}
We show an example $\logtau C_s$ in Fig.~\ref{fig:error_strings}(a) and a corresponding $P_\varsigma$ obtained via closed loops in Fig.~\ref{fig:error_strings}(b).

The normalized trace $\trn [ \logtau C_s P_\varsigma \Pi_0 ] = 1$ when $[\logtau  C_s, P_\varsigma]_- = 0$ and $\trn [ \logtau  C_s P_\varsigma \Pi_0 ] = -1$ when $[\logtau C_s, P_\varsigma]_+ = 0$ where $[A,B]_\pm = AB\pm BA$ is the (anti)commutator.
The Pauli strings $\logtau C_s$ and $P_\varsigma$ anticommute when an odd number of vertices $v$ has $S_v^X=-1$ eigenvalue in the syndrome $s$ and is simultaneously contained (i.e., $n_v=1$) in the closed-loop configuration $\logtau C_s P_\varsigma = \pm \prod_{p,v} (S_p^Z)^{\tilde{n}_p} (S_v^X)^{n_v}$; cf.\ Fig.~\ref{fig:error_strings} for an example of anticommuting Pauli strings $\logtau C_s$ and $P_\varsigma$~\footnote{We impose the convention that $P_\varsigma$ is ordered as $X\dots X Z\dots Z$ [Eq.~\eqref{eq:local_unitary}], and $\logtau C_s$ as $Z\dots Z X \dots X$, [Eq.~\eqref{eq:reference_string}].}.
This implies $\trn [ \logtau  C_s P_\varsigma \Pi_0 ] = \prod_j [\eta_{\mu,s}^{z,j}]^{(1-\sigma_{v_j} \sigma_{v_j'})/2}$ since the above condition for anticommutation is equivalent to having an odd number of sites with $\eta_{\mu,s}^{z,j}=-1$ (contained in the reference string) that neighbor an odd number of vertices with $n_v=1$ (and hence $\sigma_v = -1$).
We absorb this sign of the trace into the Pauli coefficients by updating
$J_j^0 \to J_j^0 + (i\pi/4) (1-\eta_{\mu,s}^{z,j})$ and $J_j^x\eta_{\mu,s}^{x,j} \to J_j^x\eta_{\mu,s}^{x,j} - (i\pi/4) (1-\eta_{\mu,s}^{z,j})$.

Accordingly the coefficients can be cast as a partition function $\mathcal{Z}_{\mu,s} = \sum_{\{\sigma_{v},\tilde{\sigma}_p \}} e^{-H_{\mu,s} (\{\sigma_{v},\tilde{\sigma}_p \} )}$ with
\begin{align}
 H_{\mu,s} =&-  \sum_{j} \left( J_{j}^0 + J_{j}^x \eta_{\mu,s}^{x,j} \sigma_{v_j} \sigma_{v'_j} + J_j^z \eta_{\mu,s}^{z,j} \tilde{\sigma}_{p_j} \tilde{\sigma}_{p'_j} \right.\nonumber \\
 & \left.  + J_j^y \eta_{\mu,s}^{x,j}\eta_{\mu,s}^{z,j} \sigma_{v_j} \sigma_{v'_j} \tilde{\sigma}_{p_j} \tilde{\sigma}_{p'_j} \right),
 \label{eq:hamiltonian}
\end{align}
i.e., a classical RBIM with complex couplings $J_j^x$ between spins $\sigma_{v}$ on the direct and $J_j^z$ between spins $\tilde{\sigma}_{p}$ on the dual lattice, coupled by a complex interaction $J_j^y$.
This is a disordered complex-coupling variant of the Ashkin-Teller model~\cite{Ashkin:1943jd,Baxter:1971fk}.

The form of Eq.~\eqref{eq:hamiltonian} is similar to the effective Hamiltonian that arises for incoherent errors~\cite{Bombin:2012km,Chubb:2021cn}.
To illustrate this, consider the incoherent Pauli twirl approximation described by the error channel $\mathcal{E}^\inc =\bigotimes_j \mathcal{E}_j^\inc$, Eq.~\eqref{eq:pauli_twirl}. 
The post-measurement and post-correction state is $\rho_{\mu,s}^\inc = \logtau D_s^\inc [\rho] \logtau$, where the projected error channel
\begin{equation}
  D_s^\inc [\rho] = \Pi_0 C_s \mathcal{E}^\inc [\rho] C_s^\dagger \Pi_0 = \sum_{\mu=0}^3 \mathcal{Z}_{\mu,s}^\inc \logtau \rho \logtau
  \label{eq:Ds_Pauli_channel}
\end{equation}
acts incoherently in the logical subspace.
Each coefficient $\mathcal{Z}_{\mu,s}^\inc$ equals the probability of an error string $\logtau C_s$ and equivalent strings, which is a sum of probabilities of error strings deformed by closed loops.
This is a major distinction from coherent errors where $\mathcal{Z}_{\mu,s}$ are sums of amplitudes that interfere.
Expressed as a partition function we have $\mathcal{Z}_{\mu,s}^\inc = \sum_{\{\sigma_{v},\tilde{\sigma}_p \}} e^{-H_{\mu,s} (\{\sigma_{v},\tilde{\sigma}_p \} )}$  with $H_{\mu,s}$ from Eq.~\eqref{eq:hamiltonian} but with real couplings~\cite{Chubb:2021cn} that we give for completeness in Appendix~\ref{sec:pauli_twirl}.
For such incoherent errors, the $\mathcal{Z}_{\mu,s}^\inc$ is the partition function of a classical system whose Boltzmann weights correspond to error string probabilities and where the relation between couplings and temperature~\footnote{In our notation, we absorb all temperatures in the couplings.} is fixed via a Nishimori condition~\cite{Nishimori:1981dw,Dennis:2002ds,Chubb:2021cn}.
The relation between $\mathcal{Z}_{\mu,s}$ and syndrome probabilities establishes a Nishimori condition for coherent errors.

\section{Quantum Circuit from the Transfer Matrix}
\label{sec:quantum_circuit}

For coherent errors, the coefficients $\mathcal{Z}_{\mu,s}$ and the Hamiltonian $H_{\mu,s}$ are generally complex.
This implies that, unlike the partition function that arises for incoherent errors, the coherent-error partition function does not correspond to a classical system with positive Boltzmann weights.
Accordingly, we cannot use established numerical methods to evaluate the partition function such as classical Monte Carlo simulations~\cite{Krauth:1998gt,Bombin:2012km}.
Instead, we express the partition function $\mathcal{Z}_{\mu,s}$ using a transfer matrix~\cite{Schultz:1964fv,Merz:2002gj} for the interacting Ising model.
The transfer matrix is a (1+1)D quantum circuit with nonunitary gates~\cite{Behrends:2024bs}.

We first introduce horizontal and vertical qubit labels $j\to (l,m)$; cf.\ Fig.~\ref{fig:layout}(a):
Spins labeled by even $m$ are on the direct, odd $m$ on the dual lattice; horizontal bonds on the direct lattice (and vertical bond on its dual) are labeled by even $m$, and vertical bonds on the direct lattice (and horizontal bonds on its dual) by odd $m$.
The complex partition function
\begin{align}
 \mathcal{Z}_{\mu,s} = \braket{\phi_0 | \calM | \phi_0} , & & \calM = \hat{V}^{(L)}_{\mu,s} \hat{H}^{(L-1)}_{\mu,s} \dots  \hat{V}^{(1)}_{\mu,s}
 \label{eq:circuit}
\end{align}
where $\ket{\phi_0}$ encodes the boundary conditions and $\calM$ is a quantum circuit, consisting of layers $\hat{H}^{(l)}_{\mu,s}$ and $\hat{V}^{(l)}_{\mu,s}$ for slices of horizontal and vertical couplings of the direct lattice, respectively.

Each horizontal layer consists of $M-1$ many-body gates $\hat{H}^{(l)}_{\mu,s} = \prod_{m=1}^{M-1} \Hgate$ with
\begin{equation}
 \Hgate = e^{\kappa_{lm}^{(0)} + \kappa_{lm}^{(1)} \sigma^z_{2m-1} \sigma^z_{2m+1} + \kappa_{lm}^{(2)} \sigma^x_{2m} + \kappa_{lm}^{(3)} \sigma^z_{2m-1} \sigma^x_{2m} \sigma^z_{2m+1}} ,
 \label{eq:Hlm_matrices}
\end{equation}
where the Pauli $\sigma^x_m$ and $\sigma^z_m$ act on $(2M-1)$-site 1D transfer matrix states $\ket{ \{\sigma_m\} }$, i.e., in a $2^{2M-1}$-dimensional Hilbert space.
(Recall, $M$ is the length of the shortest path for $\logZ$, see Fig.~\ref{fig:layout}(c).)
Each vertical layer consists of $M$ gates $\hat{V}_{\mu,s}^{(l)} = \prod_{m=1}^{M} \Vgate$ with, for bulk $1<m<M$,
\begin{equation}
 \Vgate = e^{\lambda_{lm}^{(0)} + \lambda_{lm}^{(1)} \sigma^z_{2m-2} \sigma^z_{2m} + \lambda_{lm}^{(2)} \sigma^x_{2m-1} + \lambda_{lm}^{(3)} \sigma^z_{2m-2} \sigma^x_{2m-1} \sigma^z_{2m}}.
 \label{eq:Vlm_matrices}
\end{equation}
We give the couplings $\kappa^{(\nu)}_{lm}$ and $\lambda^{(\nu)}_{lm}$ and more details on the boundary conditions in Appendix~\ref{sec:boundaries}.
The main feature at the top and bottom boundaries [in terms of Fig.~\ref{fig:layout}(c)] is that $\hat{V}_{\mu,s}^{(l,1)}$ and $\hat{V}_{\mu,s}^{(l,M)}$ contain single $\sigma^z_{2}$ and $\sigma^z_{2M-2}$ operators, which correspond to magnetic fields that act only at the boundaries.

To encode smooth edges at left and right sides [cf.\ Fig.~\ref{fig:layout}(c)], we use the boundary state
\begin{equation}
 \ket{\phi_0} = \sqrt{2}^{M-1} \ket{1}_1 \otimes \ket{+}_2 \otimes \dots \ket{+}_{2M-2} \otimes \ket{1}_{2M-1} ,
 \label{eq:boundary1}
\end{equation}
with spins on the dual lattice (odd $m$) initialized in the eigenstate $\sigma_{m}^z \ket{1}_m = \ket{1}_m$, and sites on the direct lattice (even $m$) in the eigenstate $\sigma_{m}^x \ket{+}_m = \ket{+}_m$.
We show a fully labeled sketch of the quantum circuit and its boundary states for $L=3$ and $M=4$ in Fig.~\ref{fig:circuit}(a).

\subsection{QEC Phase from Spontaneous Symmetry Breaking}

We now discuss the imprint of the phases of QEC on the 1D states in the transfer matrix space.
To this end, we use the quantum circuit [Eq.~\eqref{eq:circuit}] to define a 1D Hamiltonian $\calH$ via the thermal density matrix $\calM  \calM^\dagger = \exp (-L \calH)$ with inverse temperature $L$~\cite{Venn:2023fp}.
The 1D Hamiltonian $\calH$ possesses an exact and an approximate $\mathbb{Z}_2$ symmetry: The exact symmetry is given by the operator $\odd = \prod_{m=1}^{M} \sigma_{2m-1}^x$, which commutes with $\calM$ and hence with $\calH$. The approximate symmetry is given by $\even = \prod_{m=1}^{M-1} \sigma_{2m}^x$, which commutes with all $\Hgate$ and all bulk $\Vgate$, but not with $\hat{V}_{\mu,s}^{(l,1)}$ and $\hat{V}_{\mu,s}^{(l,M)}$ at the top and bottom boundaries [in terms of Fig.~\ref{fig:layout}(c)].

In the error-correcting phase, $\calH$ must be gapped and must display spontaneous breaking (SSB) of $\odd$ ~\cite{Venn:2023fp,Behrends:2024bs}, as we show by taking cues from the scenario of rotations with a small $Y$ component:
We relate the gap to $\calZ$ and $\mathcal{Z}_{\mu\oplus \logX,s}$ (where $\mu \oplus \logX$ denotes changing $\logtau$ to $\logtau \logX$) differing by a factor $\propto M e^{-L/(2\xi_0)}$ that decays with the length $L$ of $\logX$.
(The decay length $\xi_0$ is the inverse of the energy cost of a domain wall in the even-$m$ sector.)
SSB implies that $\calH$ has two lowest-energy states with opposite $\odd$ eigenvalues and exponentially small energy splitting.
We will use this to show that $\calZ$ and $\mathcal{Z}_{\mu\oplus \logZ,s}$ differ by a factor $L e^{-M/(2\xi_1)}$ that decays with the length $M$ of $\logZ$.
(The decay length $\xi_1$ is the inverse of the gap in the odd-$m$ sector.)
These features show that there is a unique largest $\calZ$ ($\mu\in \{0,1,2,3\}$) that is exponentially larger in code distance than the other $\mathcal{Z}_{\nu\neq\mu,s}$, hence the scaling of the minimum infidelity,
\begin{equation}
 \min_\mu r_{\mu,s} \propto \exp \left[ -\min\left( \frac{L}{\xi_0},\frac{M}{\xi_1} , \frac{ L+M}{\xi_2} \right) \right],
\end{equation}
is dominated by the largest of the exponentially decaying terms, where $\xi_2$ is the decay length for $Y$ errors.

We first consider the ratio of $\calZ$ and $\mathcal{Z}_{\mu\oplus \logX,s}$.
The terms $\hat{V}_{\mu,s}^{(l,1)}$ and $\hat{V}_{\mu,s}^{(l,M)}$ at the top and bottom boundaries [in terms of Fig.~\ref{fig:layout}(c)] contribute to $\calH$ as boundary magnetic fields $\propto \sigma_2^z$ and $\propto \sigma_{M-2}^z$, which for simplicity we assume contribute with the same sign.
(When the sign is opposite, the roles of $\calZ$ and $\mathcal{Z}_{\mu\oplus \logX,s}$ in the following argument are reversed.)
Both lowest-energy states are thus spin polarized.
Applying $\logX$ corresponds to flipping Ising spins in $\calZ$ along the whole length of the 2D Ising system, and hence to flipping one edge magnetic field in $\hat{\mathcal{H}}_{\mu\oplus \logX,s}$.
This increases the lowest energies of $\hat{\mathcal{H}}_{\mu\oplus \logX,s}$ by the cost of one domain wall on sites with even $m$ in $\calH$, which is, in the limit where even $m$ and odd $m$ are decoupled, the size of the gap in the even-$m$ sector.
Since there are approximately $M$ choices for the position of the domain wall, there are approximately $M$ lowest-energy states in $\hat{\mathcal{H}}_{\mu\oplus \logX,s}$.
The ratio thus scales, assuming that the overlaps between $\ket{\phi_0}$ and one-domain-wall states are all roughly equal, $\mathcal{Z}_{\mu\oplus \logX,s}/\calZ \propto M e^{-L/(2\xi_0)}$.

We now consider the ratio of $\calZ$ and $\mathcal{Z}_{\mu\oplus \logZ,s}$.
As a consequence of SSB, the lowest-energy states (with energies $E_0$ and $E_1$) are both eigenstates of $\odd$ with opposite eigenvalues.
Their energies are split exponentially $E_1-E_0 = e^{-M/\xi_1}$ in $M$ when the $\calH$ is in its ordered phase on odd sites.
The state $\ket{\phi_0}$ encoding the left and right boundaries is in a superposition of both $\odd$ sectors; cf.\ Eq.~\eqref{eq:boundary1}.
As $\calH$ preserves $\odd$, the final $\ket{\phi_N^{\{\eta\}}}=\calM \ket{\phi_0}$ must also be a superposition of both ground states, but with different amplitudes proportional to $e^{-L E_0/2}$ and $e^{-L E_1/2}$.
This implies that, when $L E_\mathrm{gap} \gg 1$, where $E_\mathrm{gap}$ is the size of gap that separates $E_0$ and $E_1$ from the rest of the spectrum,
\begin{align}
 \calZ
 &\approx e^{-\frac{L E_0}{2}} \braket{\phi_0|\Psi_0} + e^{-\frac{L E_1}{2}} \braket{\phi_0|\Psi_1} \\
 & \propto  e^{-\frac{L E_0}{2}} + e^{-\frac{L E_1}{2}} \approx 2 e^{-\frac{L E_0}{2}},
\end{align}
where $\ket{\Psi_j}$ are the eigenstates of $\calH$ with energies $E_j$, and where we assumed $\braket{\phi_0|\Psi_0} \approx \braket{\phi_0|\Psi_1}$.
(When $\braket{\phi_0|\Psi_0} \approx -\braket{\phi_0|\Psi_1}$, the roles of $\calZ$ and $\mathcal{Z}_{\mu\oplus \logZ,s}$ in the following argument are reversed.)
Applying $\logZ$ corresponds to flipping all spins in $\calZ$ along the right boundary (or equivalent choices of $\logZ$) of the 2D Ising system, as shown in Fig.~\ref{fig:layout}(c), which gives the transfer matrix $\hat{\mathcal{M}}_{\mu\oplus \logZ,s} = \odd \calM$.
Since $\ket{\Psi_0}$ and $\ket{\Psi_1}$ must have opposite $\odd$ eigenvalues, 
\begin{align}
 \mathcal{Z}_{\mu\oplus \logZ,s}
 &\approx \pm (e^{-\frac{L E_0}{2}} \braket{\phi_0|\Psi_0} - e^{-\frac{L E_1}{2}} \braket{\phi_0|\Psi_1}) \\
 &\propto  e^{-\frac{L E_0}{2}} (1 - e^{-\frac{L (E_1-E_0)}{2}}) \approx e^{-\frac{L E_0}{2}} L e^{-\frac{M}{2\xi_1}}
\end{align}
The ratio thus scales as $\mathcal{Z}_{\mu\oplus \logZ,s}/\calZ\propto L e^{-M/(2\xi_1)}$.

A rotation component about $Y$ [$n_{j,2} \neq 0$ in Eq.~\eqref{eq:rotation}] couples sectors with even and odd $m$, and hence they cannot be treated separately.
As a consequence, the emergent length $\xi_2$ determines the ratio of $\mathcal{Z}_{\mu\oplus \logY,s}$ and $\calZ$, which decreases with $e^{-(L+M)/(2\xi_2)}$ (instead of $e^{-L/(2\xi_0) - M/(2\xi_1)}$ in the decoupled limit), which concludes our argument.

While being below threshold implies a gapped $\calH$, the converse is not true:
For incoherent errors, $\calH$ above threshold is also gapped but in a disordered phase (whose $\odd$ is unbroken)~\cite{Merz:2002gj,Venn:2023fp,Behrends:2024bs}:
This implies that inserting logical operators does not exponentially separate different $\calZ$ by changing the energies.
For coherent $X$ (or $Z$) rotations, $\calH$ above threshold is gapless~\cite{Venn:2023fp}.

\subsection{Characterization via Entanglement in Transfer Matrix Space}

The transition between different phases of QEC is reflected in $\ket{\phi_N^{\{\eta\}}}$: For a bipartition of the state between sites $m\le m'$ and $m >m'$, consider the entanglement entropy
\begin{equation}
 S_{m'} = -\tr[\omega_{m'}(\phi_N^{\{\eta\}} ) \log \omega_{m'} (\phi_N^{\{\eta\}})]
\end{equation}
with the reduced density matrix $\omega_{m'} (\phi_N^{\{\eta\}}) = \tr_{m>m'} [\ket{\phi_N^{\{\eta\}} }\bra{\phi_N^{\{\eta\}}}]$.
Ground states of gapped 1D quantum systems follow an area law~\cite{Hastings:2007bu}, hence, the entanglement entropy does not grow with the 1D system size $M$ below threshold.
When $\calH$ is gapless, the entanglement entropy of the ground state~\cite{Calabrese:2009dx} and low-energy states~\cite{Alcarez:2011gx} follows an logarithmic law, i.e., it increases as $\log M$.
The same scaling has been observed for the final 1D state emerging in the quantum circuit for $X$ (or $Z$) rotations~\cite{Behrends:2024bs}.
Although, based on these arguments, it cannot be guaranteed that $\ket{\phi_N^{\{\eta\}}}$ above threshold follows a logarithmic law as $\ket{\phi_N^{\{\eta\}}}$ is not the ground state but a superposition of low-energy states since the system is not gapped above threshold, we expect that the entanglement entropy for coherent errors above threshold grows with $M$---we expect to this to hold also for more general coherent errors, e.g., for two-qubit coherent errors that exhibit a volume law~\cite{Bao2024}.

In numerical simulations, distinguishing a slow increase of the entanglement entropy from an area law can be difficult when finite size effects are strong, which makes it difficult to identify the onset of logarithmic scaling.
Instead, one can resort to $\sigma_S$, the standard derivation of the half-system entanglement entropy with respect to different syndromes~\cite{Kjall:2014bd}:
Close to the transition between distinct entanglement regimes, entanglement fluctuations are large as the entanglement entropy might, depending on the syndrome $s$, either stay constant with $M$, or increase with $M$.
Thus, a maximum in $\sigma_S$ indicates an entanglement transition.
In Sec.~\ref{sec:numerics}, we use $\sigma_S$ to identify the entanglement transition, which corresponds to the error threshold.

Thanks to the area law below threshold, simulations using MPS~\cite{Hauschild:2018bp,Cirac:2021gx} are numerically feasible~\cite{Bravyi:2014ja,Behrends:2024bs}.
For critical systems, i.e., close at the threshold for incoherent errors, and above the threshold for incoherent errors, efficient simulations with MPS are possible since the bond dimension (which is exponential in entanglement~\cite{Schuch:2008km}) necessary to represent the state grows polynomially with $M$~\cite{Verstraete:2006ir}.

\begin{figure}
\includegraphics[scale=1]{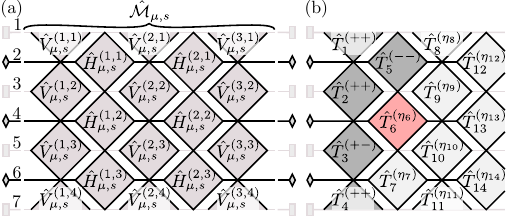}
\caption{(a) Quantum circuit $\calM$ consisting of individual gates $\Hgate$ and $\Vgate$, shown for $L=3$, $M=4$.
The boundary conditions are encoded by the product state $\ket{\phi_0}$ [Eq.~\eqref{eq:boundary1}] shown to the left and right of $\calM$, where rectangles represent $\ket{0}$ (to pick $\sigma = 1$ in the Ising model) and diamonds $\sqrt{2} \ket{+}$ (to ensure summation over $\sigma=\pm1$).
(b) Visualization of the sampling algorithm: For convenience, we label the individual gates from panel (a) by $\Tgate$; dark gray gates have a fixed error string [labeled by superscript signs $(\eta_j^x,\eta_j^z)$], light gray gates are not sampled yet, and the red gate is the currently sampled gate.
Gates are successively applied to the initial $\ket{\phi_0}$, resulting in the state $\ket{\phi_j^{\{\eta\}}}$ for the $j^\text{th}$ step (here: $j=6$), and $(\eta_j^x,\eta_j^z)$ are sampled according to the marginal distribution $P_j$ (which equals $\lVert \ket{ \phi_{j}^{\{\eta\}} }\rVert^2$ for all but the last layer of the circuit).
}
\label{fig:circuit}
\end{figure}

\section{Error String Sampling}
\label{sec:algorithm}

Since the number of syndromes grows exponentially with system size, computing exact averages $\langle \dots \rangle_s = \sum_s[\dots] P(s)$ is not feasible.
Instead, we sample syndromes $s$ from the probability distribution $P(s)$ and take the sample mean.
For incoherent errors, error strings $\logtau C_s$ can be sampled by choosing, independently for each site $j$, one of the local $\mathsf{O}_j^k$ ($k\in \{1,2,3\}$) with their respective probability $p_{j,k} = n_{j,k}^2 \sin^2 \alpha_j$ or not including an operator with probability $\cos^2 \alpha_j$ [cf.\ Eq.~\eqref{eq:pauli_twirl}].
For coherent errors, however, signs $\eta_{\mu,s}^{x/z,j}$ are generally nontrivially correlated.
While an algorithm to sample error strings is known for $X$ (or $Z$) rotations~\cite{Bravyi:2018ea}, it uses a free-fermion approach special to that case.
Hence, the same procedure cannot be applied to general rotation about any axis.
For this general case, we now introduce an algorithm to efficiently (albeit approximately) sample error strings using the quantum circuit $\calM$.

As we show, we can for each site $(l,m)$ interpret the corresponding nonunitary gate $\Vgate$ or $\Hgate$ as a weak measurements with four possible outcomes.
Apart from the last layer $\hat{V}^{(L)}_{\mu,s}$, the measurement probabilities conditioned on a 1D transfer state match the probabilities of sampling an error string with $X$ and/or $Z$ weight, conditioned on an error string encoded in the 1D transfer matrix state.

A given allocation of $\eta$ [Eq.~\eqref{eq:reference_string}] realizes an error string $\logtau C_s$, and $|\calZ|^2$ is the probability of this $\logtau C_s$ and all other strings equivalent to it (i.e., equal to it up to multiplication by stabilizers).
The configuration space of the $\eta$ thus spans all possible $\logtau C_s$, for all $s$ and $\mu \in \{0,1,2,3\}$.
Hence by using $P(\{ \eta \} ) = |\mathcal{Z}_{0,s}|^2$ and sampling from $\eta$, we sample from $P(s) = \sum_{\mu} |\calZ|^2$.
To simplify the notation for sampling from $P(\{\eta\})$, we write the quantum circuit $\hat{\mathcal{M}} = \hat{T}^{(\eta_N)}_N \dots \hat{T}^{(\eta_{1})}_{1}$ as a product of individual many-body gates $\Tgate$ with $\eta_j=(\eta^{x,j},\eta^{z,j})$ on the site of the $j^\text{th}$ physical qubit.
$\Tgate$ thus equals $\Vgate$ on vertical and $\Hgate$ on horizontal bonds of the direct lattice, where we now label the qubits by $j\in \{1,\dots N\}$ instead of by their horizontal and vertical position $(l,m)$.
We show the circuit including the many-body gates labeled by $\Tgate$ in Fig.~\ref{fig:circuit}(b).

Taking the boundary state $\ket{\phi_0}$ from Eq.~\eqref{eq:boundary1}, the error string probability is
\begin{equation}
 P ( \{\eta\} ) = \bra{\phi_0} [\hat{T}^{(\eta_1)}_1]^\dagger \dots [\hat{T}^{(\eta_N)}_N]^\dagger \omega_\phi^{(N)} \hat{T}^{(\eta_N)}_N \dots \hat{T}^{(\eta_{1})}_{1} \ket{\phi_0} 
 \label{eq:probability_circuit}
\end{equation}
with the density matrix $\omega_\phi^{(N)} = \ket{\phi_0} \bra{\phi_0}$.

We first consider the conditional probability $P(\eta_N|\eta_{N-1} \dots \eta_1)$ of $\eta_N$ for given $\eta_1 \dots \eta_{N-1}$: This is simply $P ( \{\eta\} )$ from Eq.~\eqref{eq:probability_circuit} divided by the marginal distribution $P_{N-1} (\eta_{N-1}\dots \eta_1 )$ of the first $N-1$ qubits.
The marginal distribution
\begin{equation}
 P_{N-1} ( \eta_{N-1}\dots \eta_1 ) =  \braket{\phi_{N-1}^{\{\eta\}} | \omega_\phi^{(N-1)} | \phi_{N-1}^{\{\eta\}} }
 \label{eq:marginal}
\end{equation}
is found by summing over the $\eta_N$ variable~\cite{CambridgeStat}, i.e., the density matrix $\omega_\phi^{(N-1)} = \sum_{\eta_N} [\hat{T}^{(\eta_N)}_N]^\dagger \omega_{\phi}^{(N)} \hat{T}^{(\eta_N)}_N$ is the sum over all four options for the remaining $N^\text{th}$ qubit.
The state $\ket{\phi_{N-1}^{\{\eta\}}} = \hat{T}^{(\eta_{N-1})}_{N-1} \dots \hat{T}^{(\eta_{1})}_{1} \ket{\phi_0}$ is the boundary state evolved by the first $N-1$ gates of the quantum circuit.
The conditional probability $P_{N-1} (\eta_{N-1}|\eta_{N-2} \dots \eta_1)$ of the penultimate $\eta_{N-1}$ is accordingly the marginal distribution of the first $N-1$ qubits [Eq.~\eqref{eq:marginal}] divided by the marginal distribution of the first $N-2$ qubits.
Generally, $P_{j} (\eta_{j}|\eta_{j-1} \dots \eta_1) = P_{j} (\eta_{j}\dots \eta_1)/P_{j-1} (\eta_{j-1}\dots \eta_1)$ for all $j = 1\dots N$.
The expression for the $j^\text{th}$ marginal distribution contains the density matrix
\begin{equation}
 \omega_\phi^{(j)} = \sum_{\eta_{j+1} \dots \eta_N} [\hat{T}^{(\eta_{j+1})}_{j+1} ]^\dagger \dots [\hat{T}^{(\eta_{N})}_N ]^\dagger \omega_{\phi}^{(N)} \hat{T}^{(\eta_{N})}_N \dots \hat{T}^{(\eta_{j+1})}_{j+1}
\end{equation}
which is the sum of $4^{N-j}$ terms.

We now show that we can express $\omega_\phi^{(j)}$ analytically for the boundary state corresponding to smooth edges of the direct lattice, i.e., for $\ket{\phi_0}$ [Eq.~\eqref{eq:boundary1}].
This step is crucial for the sampling algorithm that we introduce.
Starting with the density matrix
\begin{equation}
 \omega_\phi^{(N)} = \frac{1}{2^M} (\mathbb{1} +\sigma_1^z ) (\mathbb{1}+\sigma_2^x) \dots  (\mathbb{1}+\sigma_{2M-1}^z) .
\end{equation}
The operator $T_N^{(\eta_N)}$ is in the last layer~\footnote{Here the $\hat{T}_n^{(\eta_n)}$ sites are wlog ordered from left to right.}, i.e., it is of the form $\hat{V}^{(L,m)}$. Summing over all possible $\eta_N$, the Pauli matrices on all transfer matrix sites that $\hat{V}^{(L,m)}$ acts on are replaced by identities.
Choosing a qubit labeling such that the $N^\text{th}$ qubit has $\hat{T}_N^{(\eta_N)} =\hat{V}^{(L,M)}$, the density matrix relevant for the $(N-1)^\text{th}$ marginal distribution
\begin{equation}
 \omega_\phi^{(N-1)} = (\mathbb{1} +\sigma_1^z) \dots  (\mathbb{1}+\sigma_{2M-3}^z) \mathbb{1}_{2M-2} \mathbb{1}_{2M-1} ,
\end{equation}
where we introduced labels $\mathbb{1}_{m}$ for identity matrices on all sites $m$ that $\hat{T}_N^{(\eta_N)}$ has acted on.
This pattern continues until all sites are replaced by the identity and we remain with the identity matrix for those $\Tgate$ that are not in the last layer: $\omega_\phi^{(j<N-M)} = \mathbb{1}$.
All marginal distributions can thus be computed using the evolved 1D transfer matrix state and $\omega_\phi^{(j)}$, i.e., $P_{j} ( \eta_{j}\dots \eta_1 ) =  \braket{\phi_{j}^{\{\eta\}} | \omega_\phi^{(j)} | \phi_{j}^{\{\eta\}} }$ [also cf.\ Fig.~\ref{fig:circuit}(b)].
When $j<N-M$, $P_{j} =  \lVert \ket{ \phi_{j}^{\{\eta\}} }\rVert^2$, i.e., the probabilities for $\eta_j$ are proportional to the amplitudes of the evolved states $\ket{\phi_j^{\{\eta\}}}$.
In the $j^\text{th}$ step, using the marginal distribution, one can now sample $\eta_j$ based on its probability~\cite{Krauth:1998gt}.

This sampling algorithm is crucial for the simulation of generic rotation axes.
It is beyond the scope of algorithms for $X$ (or $Z$) rotations~\cite{Bravyi:2018ea} and tensor-network based methods~\cite{Darmawan:2017fx,Darmawan2024} since it can sample errors string in geometries where $P(s|\rho)\neq P(s)$.
Similar ideas can be extended to generic non-Pauli errors~\cite{Behrends2024}, greatly extending the regime of existing algorithms suitable for errors close to the incoherent limit~\cite{Ma:2023ia}.

The main requirement is the ability to compute $\ket{\phi_j^{\{\eta\}}}$ efficiently.
This can be done, approximately but to high accuracy, using MPS methods provided $\ket{\phi_j^{\{\eta\}}}$ has low entanglement.
Our methods thus allows us to chart the error-correcting phase (where $\ket{\phi_j^{\{\eta\}}}$ satisfies an area law) and  to study the behavior above threshold (with logarithmic entanglement in $\ket{\phi_j^{\{\eta\}}}$).

\section{Numerical Threshold Estimates}
\label{sec:numerics}

\begin{figure*}
\includegraphics[scale=1]{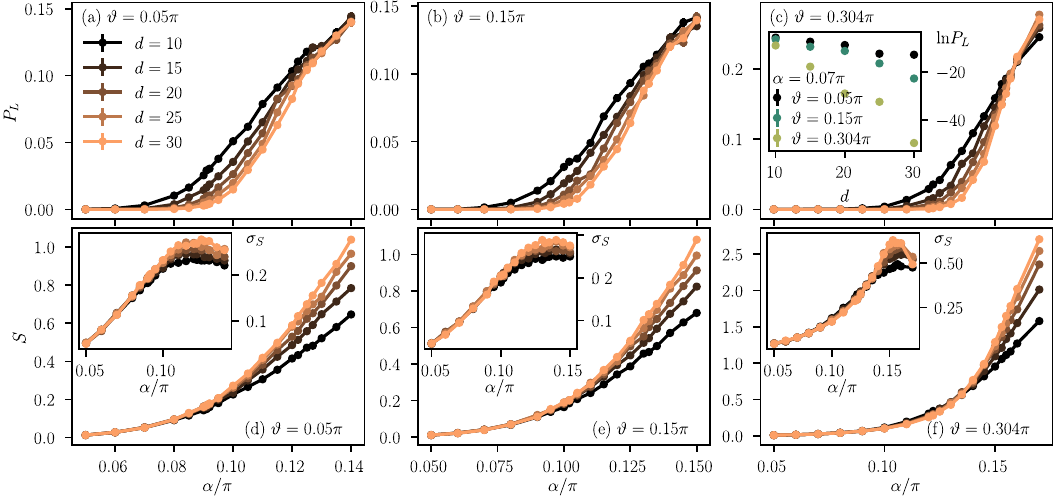}
\caption{(a)--(c) Syndrome-averaged minimum infidelity and (d)--(f) half-system entanglement entropy as a function of the coherent rotation angle $\alpha$, averaged over 1000 to 10000 error strings.
The different colors denote the code distance $d$, and the angle $\vartheta$ of the rotation axis is shown in the panels.
The inset in (c) shows the exponential decay of $P_L$ with code distance below threshold ($\alpha=0.07\pi$), where the colors denote different $\vartheta$.
Error bars indicating the standard error are imperceptible.
The insets in (d)--(f) show $\sigma_S$, the standard deviation of the entanglement entropy, whose maximum we use to identify the error threshold.}
\label{fig:minimal_rmu}
\end{figure*}

\begin{figure*}
\includegraphics[scale=1]{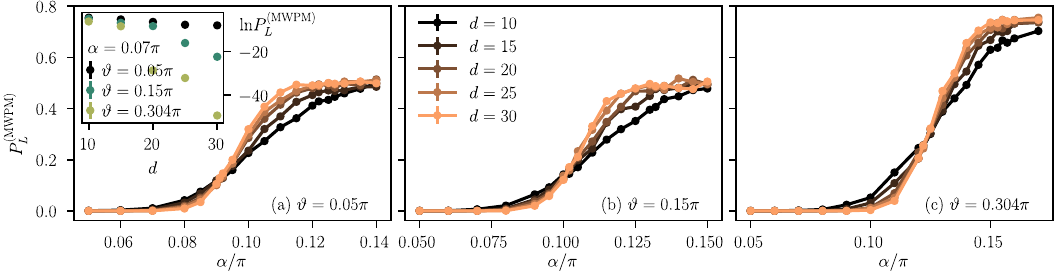}
\caption{Logical error from MWPM $P_L^\MWPM$ as a function of the coherent rotation angle $\alpha$, averaged over 1000 to 10000 error strings; error bars are imperceptible.
The different colors denote the code distance $d$, and the angle $\vartheta$ of the rotation axis is shown in the panels.
The inset in (a) shows the exponential decay of $P_L^\MWPM$ with code distance below threshold ($\alpha=0.07\pi$).}
\label{fig:mwpm}
\end{figure*}

\begin{figure*}
\includegraphics[scale=1]{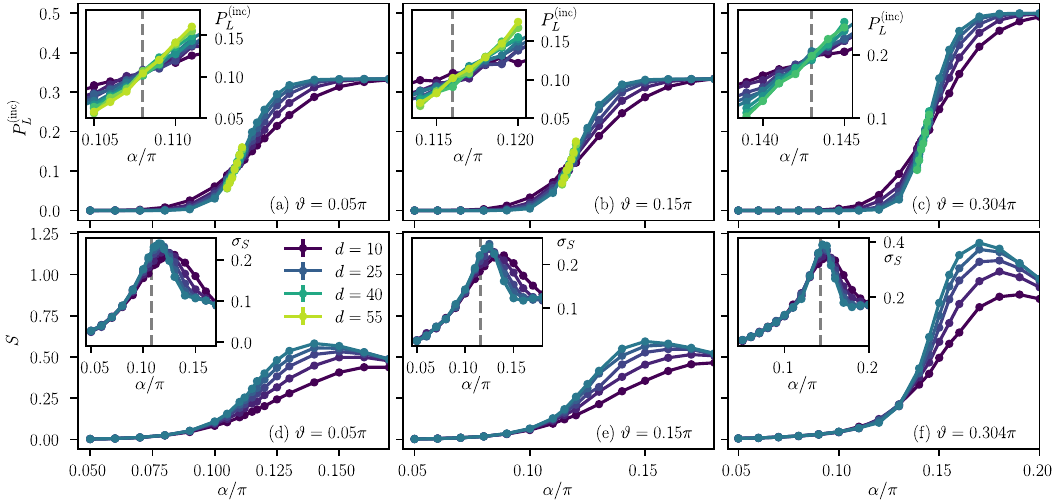}
\caption{(a)--(c) Logical error $P_L^\inc$ a function of rotation angle in the Pauli twirl approximation [Eq.~\eqref{eq:pauli_twirl}], averaged over 1000 to 10000 error strings; error bars are imperceptible.
The different colors denote the code distance $d$, and the angle $\vartheta$ of the rotation axis is shown in the panels.
The insets in (a)--(c) show $P_L^\inc$ close to the threshold marked by the gray dashed line.
(d)--(f) Bipartition entanglement entropy as a function of the rotation angle. Two area-law phases are separated by the incoherent Pauli twirl threshold.
The insets in (d)--(f) show $\sigma_S$, the standard deviation of the entanglement entropy, whose maximum indicates the threshold.}
\label{fig:incoherent}
\end{figure*}

\begin{figure*}
\includegraphics[scale=1]{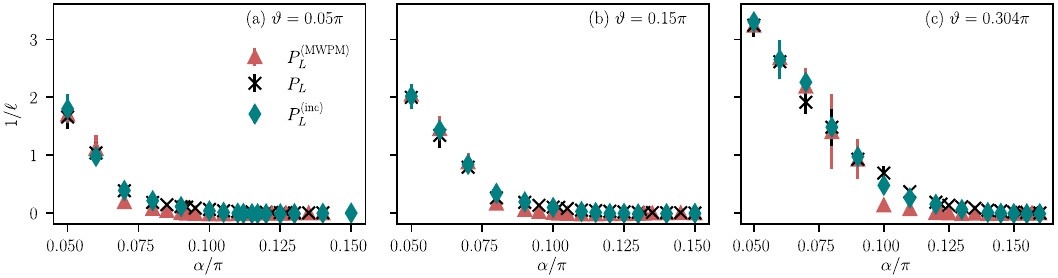}
\caption{Decay rate $1/\ell$ of the logical error as a function of the coherent rotation angle $\alpha$; error bars show the standard error of the estimated slope.
The angle $\vartheta$ of the rotation axis is shown in the panels.
The symbols and colors denote the inverse decay length of the logical error $P_L$ (black crosses), of the MWPM error $P_L^\MWPM$ (red triangles), and of the logical error $P_L^\inc$ in the incoherent Pauli twirl approximation (teal diamonds).
}
\label{fig:lengths}
\end{figure*}

We now give estimates for error thresholds.
To this end, we numerically simulate the quantum circuit introduced in Sec.~\ref{sec:quantum_circuit} for syndromes generated by the algorithm described in Sec.~\ref{sec:algorithm}.
We compute the average infidelity, entanglement entropy of the 1D transfer matrix state, and the MWPM threshold.
We compare all findings with incoherent errors in the Pauli twirl approximation.

In all simulations, we use the same rotation angles $\alpha_j \to \alpha$ and rotation axes $\mathbf{n}_j \to \mathbf{n}$ for all qubits $j$.
We consider three rotation directions $\vartheta \in \{ 0.05\pi, 0.15\pi, 0.304\pi \}$ and $\varphi = \pi/2$ as shown in Fig.~\ref{fig:layout}(a), i.e., we start with just a slight deviation from $X$ rotations ($\vartheta=0.05\pi$) and end with rotations in the $(1,1,1)$ direction ($\vartheta=0.304\pi$).
We compare coherent errors with the incoherent Pauli twirl approximation, where the $(1,1,1)$ directions corresponds to depolarizing noise (equally likely $X$, $Y$, and $Z$ errors).
In all simulations, we consider square geometries $d=M=L$ and label the system size by the code distance $d$.
The total number of qubits is $N=d^2+(d-1)^2$ and the number of $S_v^X$ and $S_p^Z$ stabilizers is $d(d-1)$ each.
For the MPS representation we use a maximal bond dimension of $\chi_\text{max} = 128$ that we found sufficient for all system sizes and angles we considered.
(The maximal entanglement entropy captured by an MPS with $\chi_\text{max}=128$ is well above the largest value we numerically observed.)

In Fig.~\ref{fig:minimal_rmu}(a)--(c), we show the logical error $P_L$ as a function of the rotation angle $\alpha$ for different distances $d$.
We identify two phases: Below a threshold angle $\alpha_\thr$, $P_L$ decreases exponentially (within error bars) with code distance; cf.\ inset in Fig.~\ref{fig:minimal_rmu}(c).
Above threshold, the minimum infidelity continues to decrease, but much more slowly, with code distance; the decrease is consistent with a power-law decrease to a positive constant.
Similar behavior has been observed previously for $X$ rotations~\cite{Bravyi:2018ea,Venn:2020ge,Venn:2023fp}, and it is distinct from the above-threshold regime for incoherent errors, where the infidelity increases with code distance~\cite{Fowler:2012fi}.
From the scaling of the minimum infidelity, we can give only a rough lower bound on the threshold $\alpha_\thr$:
We observe a clear exponential decrease with code distance for angles $\alpha <\alpha_\thr$, which gives the estimates $\alpha_\thr >0.09\pi $, $\alpha_\thr >0.1\pi $, and $\alpha_\thr >0.11\pi $ for the axes $\vartheta=0.05\pi$, $\vartheta=0.15\pi$ and $\vartheta=0.304\pi$, respectively.

To estimate $\alpha_\thr$ more accurately, we distinguish the phases of QEC by the different scaling of the half-system entanglement entropy of the evolved transfer matrix state.
Below threshold, the half system entanglement entropy $S$ follows an area law, while above threshold, it increases with code distance $d$, as we show in Fig.~\ref{fig:minimal_rmu}(d)--(f).
Numerically we find that $S$ increases approximately logarithmically with $d$.
In the insets in Fig.~\ref{fig:minimal_rmu}(d)--(f), we show the standard deviation $\sigma_S$ of the half-system entanglement entropy $S$.
Since the maximal peak position of $\sigma_S$ is a proxy for the threshold, we obtain the threshold estimates $\alpha_\thr/\pi = 0.12(1),0.13(1),0.155(5)$ for $\vartheta/\pi =0.05,0.15,0.304$, respectively.

We now identify the MWPM~\cite{Fowler:2012ji,Fowler:2015hb} scaling and threshold.
For each syndrome $s$, we use PyMatching~\cite{pymatchingv1,pymatchingv2} to select one correction operation $C_s \bar{\mathsf{O}}_{\mu^\MWPM}$ whose probability is $P^\MWPM (s) = |\mathcal{Z}_{\mu^\MWPM,s}|^2$.
We determine the MWPM threshold by computing the logical error probability $P_L^\MWPM = \langle 1 - P^\MWPM (s)/P(s) \rangle_s$.
Below threshold, this goes exponentially to zero with code distance, as we show in the inset in Fig.~\ref{fig:mwpm}(a).
Above threshold, $P_L^\MWPM$ increases with code distance until it saturates.
In Fig.~\ref{fig:mwpm}, we show $P_L^\MWPM$ as a function of rotation angle for different code distances.
From the numerical data, we estimate $\alpha_\thr^\MWPM/\pi = 0.095(5),0.10(1),0.12(1)$---significantly smaller than the maximum likelihood thresholds $\alpha_\thr$.

To gain further insight into the surface code performance for coherent errors, we compare the error threshold and scaling of the error rate with the corresponding incoherent Pauli twirl approximation.
We show the logical error $P^\inc_L$ in Fig.~\ref{fig:incoherent}(a)--(c).
Here, different from coherent errors (but similarly to their MWPM correction), a threshold clearly separates phases where the logical error rate decreases with code distance (below threshold) and increases with code distance (above threshold).
We estimate the thresholds $\alpha^\inc_\thr = 0.108(1)\pi$, $\alpha^\inc_\thr = 0.116(1)\pi$, and $\alpha^\inc_\thr = 0.143(2)\pi$ for $\vartheta = 0.05\pi$, $\vartheta =0.15\pi$ and $\vartheta =0.304\pi$, respectively.
For $\vartheta=0.304\pi$, the corresponding probability $p_\thr = \sin^2 \alpha^\inc_\thr = 0.189(4)$ is consistent with the known threshold for depolarizing noise~\cite{Bombin:2012km}.

The Pauli twirl also allows us to benchmark the use of entanglement for determining thresholds.
In Fig.~\ref{fig:incoherent}(d)--(f), we show the half-system entanglement entropy $S$ in the transfer matrix space and demonstrate that $S$ increases with the transverse dimension $M$ around the incoherent-error threshold.
Similarly to coherent errors, we can identify the threshold by the peak of the entanglement entropy standard deviation $\sigma_S$, which we show in the insets of Fig.~\ref{fig:incoherent}(d)--(f).
Note that for incoherent errors we can identify the threshold more accurately by the scaling of $P_L^\inc$ itself---the maximum of $\sigma_S$ approaches the threshold $\alpha^\inc_\thr$ with increasing system size.
For moderate system sizes, however, using $\sigma_S$ to estimate threshold values gives only a rough estimate.
This indicates that the thresholds we identified for coherent errors using the same method can be improved by investigating larger system sizes.

Finally, we compare the below-threshold behavior for coherent maximum likelihood decoding, MWPM, and the incoherent Pauli twirl approximation.
In Fig.~\ref{fig:lengths}, we show the decay rate $1/\ell$ resulting from the fitting $P_L \propto \exp (-d/\ell)$.
We compare three decay lengths $\ell$, $\ell^\MWPM$, and $\ell^\inc$, corresponding to $P_L$ (coherent maximum likelihood decoding), $P_L^\MWPM$ (coherent MWPM decoding), and $P_L^\inc$ (maximum likelihood decoding in the incoherent Pauli twirl approximation), respectively.
We find that $\ell$ and $\ell^\coh$ are comparable much below threshold, but that $P_L$ decays faster than $P_L^\coh$ close to the threshold, hence the decay rate $1/\ell$ is larger than $1/\ell^\coh$.
The MWPM decay rate $1/\ell^\MWPM$ is, within error bars, always smaller than $\ell$ and $\ell^\coh$.

\section{Conclusion and Outlook}
\label{sec:conclusion}

In this work, we developed a statistical physics mapping that describes the surface code under any single-qubit coherent error.
This mapping goes substantially beyond mappings for coherent errors with only $X$ (or $Z$) rotations, as it does not rely on a special free-fermion solubility that is present for those axes.

Using our mapping, we established error thresholds for any single-qubit coherent error in the surface code.
The error thresholds separate a quantum error-correcting regime, where the logical error decreases exponentially with code distance, from a non-correcting regime where the logical error still decreases, but much more slowly (consistently with a power law), until it saturates to a nonzero value.

The statistical physics mapping that we developed expresses error amplitudes as the partition function of a classical interacting random bond Ising model with complex couplings.
The interacting Ising model can be understood as a disordered Ashkin-Teller model with complex couplings~\cite{Ashkin:1943jd} whose real variant---which, unlike this complex-coupling model, corresponds to a classical model with positive Boltzmann weights---describes incoherent errors~\cite{Bombin:2012km,Chubb:2021cn}.

Using the transfer matrix, we expressed the partition function as a nonunitary interacting (1+1)D quantum circuit.
We found that, in the (1+1)D quantum circuit, the threshold separates an area law from an approximately logarithmic law regime of the 1D transfer matrix states.
It is this area law, and also the moderate entanglement in the logarithmic phase, that we could use to simulate the system using matrix product states.

The logarithmic phase above threshold, where the logical error rate $P_L$ decreases to a nonzero value, is distinct from above-threshold phase for incoherent errors with area-law entanglement~\cite{Behrends:2024bs} and increasing $P_L$.
The latter corresponds to a disordered phase of the statistical mechanics model~\cite{Dennis:2002ds,Merz:2002gj}.
In the coherent case, based on the behavior consistent with a power-law decay of $P_L$, and on the logarithmic entanglement, we interpret this phase as displaying quasi-long-range order:
This can be thought of of as an interacting form of the ``metallic'' phase found for $X$ (or $Z$) rotations~\cite{Venn:2023fp,Behrends:2024bs}.
The logarithmic phase is also distinct from the volume-law phase found in the above-threshold quantum circuit for two-qubit coherent errors in the surface code~\cite{Bao2024}, and for generic random unitary quantum circuits~\cite{Calabrese:2005dy}.

The error rates, thresholds, and entanglement features are all understood in the sense of a syndrome average---the statistical approach required to assess the performance of QEC under the Born rule.
To compute such averages, we introduced an algorithm based on our quantum circuit that draws errors according to their Bloch-sphere averaged probability.
Since the implementation of this algorithm relies on an MPS approximation, in practice the errors are sampled according to their approximate probability $P'(s)$, which is close but not equal to the syndrome probability $P(s)$ due to the discarded weight in the truncated MPS states~\cite{Hauschild:2018bp}.
This approximation is similar to tensor-network-based approaches for syndrome sampling~\cite{Darmawan:2017fx,Darmawan2024}, which typically employ MPS for tensor contraction~\cite{Darmawan:2017fx}.
Different from tensor-network sampling, our algorithm samples directly from $P(s)$ with respect to a suitable average over encoded states $\rho$. This is advantageous for systems where the syndrome probability for a given $\rho$ depends on $\rho$, as is the case for surface codes with odd-weight stabilizers~\cite{Venn:2020ge}.

Future research directions include the generalization of our approach to other quantum codes under coherent errors.
As the MPS-based simulation of the quantum circuit representing the partition function does not rely on a free-fermion solubility, but requires only that the code be local, it  should be adaptable to a broad family of quantum codes including the color code~\cite{Bombin:2006hw}, or codes beyond qubits such as $\mathbb{Z}_n$ surface codes, $\mathbb{Z}_n$ color codes, the double-semion code~\cite{Levin:2005fy,Dauphinais:2019bl}, etc.

Our work could also be extended towards other non-Pauli errors, e.g., the amplitude damping channel, or Clifford errors and their deformations~\cite{Magesan:2013jc,Gutierrez:2013ei}.
We expect that these errors also admit a representation in terms of a suitable statistical-mechanics model with complex couplings.
Our framework could also be extended to multi-qubit errors, albeit we anticipate that these will require a more complicated sampling algorithm.

Since, as we have shown, the fundamental threshold for maximum likelihood decoding is larger than the minimum weight perfect matching (MWPM) threshold, our work could inspire decoding algorithms that incorporate coherent errors better and reach thresholds closer to the maximum-likelihood bound.

\paragraph*{Note added.} Readers may also be interested in two related independent works, one on coherent errors whose angles are drawn from a random distribution~\cite{Lee2024}, and one on coherent one- and two-qubit errors~\cite{Bao2024}.

\begin{acknowledgments}
We thank Alaric Sanders for helpful discussions, and Florian Venn for previous collaborations on coherent errors.
This work was supported by EPSRC Grant No.\ EP/V062654/1, a Leverhulme Early Career Fellowship and the Newton Trust of the University of Cambridge.
Our simulations used resources at the Cambridge Service for Data Driven Discovery operated by the University of Cambridge Research Computing Service (\href{https://www.csd3.cam.ac.uk}{www.csd3.cam.ac.uk}), provided by Dell EMC and Intel using EPSRC Tier-2 funding via grant EP/T022159/1, and STFC DiRAC funding (\href{https://www.dirac.ac.uk}{www.dirac.ac.uk}).
\end{acknowledgments}

\appendix

\section{Average infidelity}
\label{sec:infidelity}

In the main text, we show that the average infidelity can be computed using only absolute values of complex partition functions, Eq.~\eqref{eq:average_infidelity}.
Here we provide more details on the infidelity and its initial-state dependence.
Taking the initial state $\rho=\ket{\psi} \bra{\psi}$ with $\ket{\psi} = \cos (\theta/2) \ket{0}_L + \sin (\theta/2) e^{i \phi} \ket{1}_L$ and the logical states $\ket{0}_L$ and $\ket{1}_L$, the fidelity of $\rho$ and the post-error, post-measurement, and post-correction state $\rho_{\mu,s} = \logtau D_s \rho D_s^\dagger \logtau$ is
\begin{equation}
 F(\rho, \rho_{\mu,s} ) = \frac{\braket{\psi| \rho_{\mu,s} | \psi}}{\tr[\rho_{\mu,s}]} = \frac{| \braket{\psi| \logtau D_s |\psi}|^2}{P(s|\rho)} .
\end{equation}
Its Bloch-sphere average equals
\begin{align}
\langle F(\rho, \rho_{\mu,s}) \rangle_\Omega =& \int_\Omega d\rho P(\rho|s) F(\rho, \rho_{\mu,s}) \\
 =& \frac{1}{P(s)} \int_\Omega d\rho  P(\rho)| \braket{\psi | \logtau  D_s |\psi }|^2 \nonumber ,
\end{align}
where we used $P(\rho|s) = P(s|\rho) P(\rho)/P(s)$ to simplify above integrand.
For uniformly distributed $P(\rho)$, the average $\int_\Omega d\rho  P(\rho) = (4\pi)^{-1}\int d\phi d\theta \sin\theta$, and the fidelity
\begin{align}
  \langle F(\rho, \rho_{\mu,s}) \rangle_\Omega
  &= \frac{1}{4\pi P(s)} \int d\phi d\theta \sin\theta | \braket{\psi | \logtau D_s |\psi }|^2 \nonumber \\
  &= \frac{1}{P(s)} \left( |\mathcal{Z}_{\mu,s}|^2 + \frac{1}{3} \sum_{\nu\neq \mu} |\mathcal{Z}_{\nu,s}|^2 \right) .
\end{align}
The conditional syndrome probability
\begin{align}
 P & (s|\rho) = 2 \cos \theta \left[ \Re( \mathcal{Z}_{0,s}^* \mathcal{Z}_{3,s}) + \Im (\mathcal{Z}_{2,s}^* \mathcal{Z}_{1,s} )\right] \\
 &+ 2\sin \theta \left\{ \cos \phi \left[ \Re( \mathcal{Z}_{0,s}^* \mathcal{Z}_{1,s} ) + \Im ( \mathcal{Z}_{3,s}^* \mathcal{Z}_{2,s} ) \right] \right. \nonumber \\
 & \left. + \sin \phi \left[ \Re( \mathcal{Z}_{0,s}^* \mathcal{Z}_{2,s})  + \Im (\mathcal{Z}_{1,s}^* \mathcal{Z}_{3,s}) \right] \right\} + \sum_{\mu=0}^3 |\mathcal{Z}_{\mu,s}|^2 \nonumber 
\end{align}
averaged over the Bloch sphere gives $P(s) = (4\pi)^{-1} \int_\Omega d\rho P(s|\rho) = \sum_\mu |\mathcal{Z}_{\mu,s}|^2$.
Thus, the average infidelity equals
\begin{equation}
  r_{\mu,s} = \langle 1- F(\rho, \rho_{\mu,s}) \rangle_\Omega
  = \frac{2}{3} \frac{\sum_{\nu\neq \mu} |\mathcal{Z}_{\nu,s}|^2 }{P(s)} ,
\end{equation}
as given in the main text, Eq.~\eqref{eq:average_infidelity}.

\section{Details on Pauli twirl approximation}
\label{sec:pauli_twirl}

In the Pauli twirl approximation, the error channel projected onto the logical subspace acts incoherently; cf.\ Eq.~\eqref{eq:Ds_Pauli_channel}.
The probabilities $\mathcal{Z}_{\mu,s}^\inc = \sum_{\{\sigma_v,\tilde{\sigma}_p\}} \exp (H_{\mu,s}^\inc)$ can be expressed as classical partition functions with the Hamiltonian given in Eq.~\eqref{eq:hamiltonian}.
Its real couplings~\cite{Chubb:2021cn}
\begin{subequations}
\begin{align}
 J_j^{0,\inc} =& \frac{1}{2} \log \left| n_{j,1} n_{j,2} n_{j,3} \cos \alpha_j \sin^3 \alpha_j \right| \\
 J_j^{x,\inc} =& \frac{1}{2} \log \left| \frac{n_{j,1} \cot\alpha_j }{n_{j,2} n_{j,3}} \right| \\
 J_j^{y,\inc} =& \frac{1}{2} \log \left| \frac{n_{j,2} \cot\alpha_j }{n_{j,1} n_{j,3}} \right| \\
 J_j^{z,\inc} =& \frac{1}{2} \log \left| \frac{n_{j,3} \cot\alpha_j }{n_{j,1} n_{j,2}} \right|
\end{align}\label{eq:real_couplings}\end{subequations}
give positive Boltzmann weights $\exp (H_{\mu,s}^\inc)$, each corresponding to the probability of an error chain~\cite{Dennis:2002ds}.

\section{Boundary conditions and transfer matrix details}
\label{sec:boundaries}

In this appendix we provide more details on the boundary conditions we used and its implications on the transfer matrix formulation.
Throughout this work, we use the planar surface code whose left and right edges are smooth and whose top and bottom edges are rough, as shown in Fig.~\ref{fig:layout}(c).
When expressing the complex partition function $\mathcal{Z}_{\mu,s}$ as a quantum circuit, we introduce transfer matrices $\hat{H}^{(l)}_{\mu,s}$ and $\hat{V}^{(l)}_{\mu,s}$ for the horizontal and vertical bonds of the direct lattice, respectively.
The layer $\hat{H}^{(l)}_{\mu,s}$ must fulfill
\begin{equation}
 \braket{ \{ \sigma_{m} \} | \hat{H}_{\mu,s}^{(l)} | \{ \sigma_m' \} }
 =e^{\sum_{m=1}^{M-1} h_{lm}} \delta_{ \{ \sigma_{2m+1} \}, \{ \sigma_{2m+1}' \} }
\end{equation}
with
\begin{align}
 h_{lm} =& J_{l,2m}^x \eta_{l,2m}^{x} \sigma_{2m} \sigma_{2m}' + J_{l,2m}^z \eta_{l,2m}^{z} \sigma_{2m-1} \sigma_{2m+1} \nonumber \\
 &+ J_{l,2m}^y \eta_{l,2m}^{x} \eta_{l,2m}^{z} \sigma_{2m} \sigma_{2m}' \sigma_{2m-1} \sigma_{2m+1} + J_{l,2m}^0 \nonumber  ,
\end{align}
which is satisfied by Eq.~\eqref{eq:Hlm_matrices} with transfer matrix couplings
\begin{align}
\kappa_{lm}^{(0)} &= \frac{\log \left[ \sin^2 \alpha \sin^2 \vartheta \left(1 -\sin^2 \alpha \sin^2 \vartheta\right)\right]+ i \pi  \eta_{l,2m}^x }{4} \nonumber \\
\kappa_{lm}^{(1)} &= \eta_{l,2m}^z \left(\frac{1}{4} \log \left(\frac{1-\sin^2 \alpha \sin^2 \vartheta  }{\sin ^2 \alpha  \sin ^2 \vartheta }\right)-\frac{i\pi}{4}\right) \nonumber \\
\kappa_{lm}^{(2)} &= \frac{i}{2} \left(\arctan( \tan \alpha \cos \vartheta ) +\varphi \right)+\frac{1}{4} \pi  i (1-\eta_{l,2m}^x) \nonumber \\
\kappa_{lm}^{(3)} &= \frac{i}{2} \eta_{l,2m}^z \left(\arctan (\tan \alpha \cos \vartheta )-\varphi \right) .
\end{align}
Similarly, the layer $\hat{V}^{(l)}_{\mu,s}$ encoding the vertical bonds of the direct lattice must fulfill
\begin{equation}
 \braket{ \{ \sigma_{m} \} | \hat{V}_{\mu,s}^{(l)} | \{ \sigma_m' \} }
 =e^{\sum_{m=1}^{M} v_{lm}} \delta_{ \{ \sigma_{2m} \}, \{ \sigma_{2m}' \} }
\end{equation}
\begin{widetext}
\noindent with
\begin{align}
 v_{lm} =& J_{l,2m-1}^x \eta_{l,2m-1}^{x} \sigma_{2m-2} \sigma_{2m} + J_{l,2m-1}^z \eta_{l,2m-1}^{z} \sigma_{2m-1} \sigma_{2m-1}' + J_{l,2m-1}^y \eta_{l,2m-1}^{x} \eta_{l,2m-1}^{z} \sigma_{2m-2} \sigma_{2m} \sigma_{2m-1} \sigma_{2m-1}' \nonumber \\
 &+ J_{l,2m-1}^0 ,
\end{align}
which for $m>1$ and $m<M$ is satisfied by Eq.~\eqref{eq:Vlm_matrices} with transfer matrix couplings
\begin{align}
\lambda_{lm}^{(0)} &= \frac{1}{4} \log \left[ \sin^2 \alpha  \left(1-\cos^2 \varphi  \sin^2 \vartheta \right) \left( 1-\sin^2 \alpha \left(1-\cos^2 \varphi \sin^2 \vartheta \right)\right)\right]+\frac{i \pi  \eta^z_{l,2m-1}}{4} \\
\lambda_{lm}^{(1)} &= \eta_{l,2m-1}^x \left(\frac{1}{4} \log \left(\frac{1-\sin^2 \alpha \left(1-\cos^2 \varphi \sin^2 \vartheta \right)}{\sin^2 \alpha \left(1-\cos^2 \varphi \sin^2 \vartheta \right)}\right)-\frac{i\pi}{4} \right) \\
\lambda_{lm}^{(2)} &= \frac{1}{2} i \left[ \arctan( \tan \alpha  \cos \varphi \sin \vartheta ) + \atan(\cos \vartheta ,\sin \varphi \sin \vartheta )\right] + \frac{1}{4} \pi  i (1-\eta_{l,2m-1}^z) \\
\lambda_{lm}^{(3)} &= \frac{1}{2} i \eta_{l,2m-1}^x \left[ \arctan (\tan \alpha \cos \varphi \sin \vartheta )-\atan(\cos \vartheta ,\sin \varphi \sin \vartheta )\right] ,
\end{align}
\end{widetext}
where $\atan (y,x)$ is the two-argument arctangent.
The boundary terms that encode the rough edges at top and bottom are
\begin{align}
 \hat{V}_{\mu,s}^{(l,1)} &= e^{\lambda_{l1}^{(0)} + \lambda_{l1}^{(1)} \sigma^z_{2} +    \lambda_{l1}^{(2)} \sigma^x_{1}    + \lambda_{l1}^{(3)} \sigma^x_{1} \sigma^z_{2} } \\
 \hat{V}_{\mu,s}^{(l,M)} &= e^{\lambda_{lM}^{(0)} + \lambda_{lM}^{(1)} \sigma^z_{2M-2} + \lambda_{lM}^{(2)} \sigma^x_{2M-1} + \lambda_{lM}^{(3)} \sigma^z_{2M-2} \sigma^x_{2M-1} } \nonumber .
\end{align}

\bibliography{references}

\end{document}